\documentclass[12pt]{article}

\usepackage{graphicx}
\usepackage{amsmath, physics, amsfonts, amssymb}
\usepackage{multicol}
\usepackage{biblatex} % Imports biblatex package
\usepackage{authblk}
\usepackage{dcolumn}
\usepackage{bm}
\usepackage{comment}
\usepackage{color}

\begin{document}

\title{Inflaton Dynamics in Higher-Derivative Scalar-Tensor Theories of Gravity}

\author[1]{Sam E. Brady\footnote{Corresponding author --- s.brady@qmul.ac.uk}}
\author[1]{Katy Clough}
\author[1]{Pau Figueras}
\author[1]{\'Aron D. Kov\'acs}

\affil[1]{Centre for Geometry, Analysis and Gravitation, School of Mathematical Sciences, Queen Mary University of London, Mile End Road, London E1 4NS, United Kingdom}

\date{\today}

\maketitle

\begin{abstract}
During inflation, higher-derivative terms in the gravitational action may play a significant role. Building on new stable formulations of four-derivative scalar-tensor theories, we study the impact of these corrections in the case where the inflaton is also the additional scalar degree of freedom of the modified theory.  
This case is highly restricted by requiring that in the homogeneous limit inflation must still work, and that the initial data must be in the weak coupling limit to respect the validity of the effective theory. In such cases, the non-linear dynamics of large perturbations are very similar to the GR case, with the main deviations captured by the terms relating to the homogeneous Einstein-scalar-Gauss-Bonnet contributions. We show that in principle it is possible to dynamically drive the field out of the weak-coupling regime from a starting point well within it, but that to do so one has to finely tune the setup, so such cases are unlikely to occur generically. This work provides a basis for the study of less restricted models in future, for example those in which the inflaton and scalar degree of freedom are independent, or in which one is not restricted to a weakly coupled regime.

\end{abstract}

\newpage

%\begin{multicols}{2}

\section{\label{sec:level1}Introduction}

The theory of cosmic inflation \cite{Guth:1980zm, Linde:1981mu, Albrecht:1982wi, Starobinsky:1980te,Linde:1983gd} was proposed to solve various problems in the standard model of cosmology. One of the main observations that inflation proposes to address is the high degree of homogeneity in the cosmic microwave background. Without some process to drive the universe into a highly homogeneous state, the initial conditions for cosmic evolution would require homogeneity across $\sim10^{83}$ causally disconnected regions \cite{Guth:1980zm}. Inflation addresses this by proposing a period of accelerated expansion, or equivalently a decreasing comoving Hubble radius. This can be achieved by a real scalar field slowly-rolling down a roughly flat section of its potential, such that the potential contribution to the energy density dominates over the kinetic terms.\footnote{An alternative process is a period of slow contraction followed by a bounce, see \cite{Brandenberger:2016vhg,Ijjas:2018qbo} for reviews.}

For the mechanism of inflation to consistently solve the homogeneity problem, it must itself be robust against large spatial inhomogeneities in the initial data. Inflation is known to be robust against small perturbations for a range of valid models, but the non-perturbative case is still being investigated. Building on earlier works \cite{Goldwirth:1989pr,Goldwirth:1989vz,Goldwirth:1991rj,Laguna:1991zs,Kurki-Suonio:1993lzy}, in \cite{East:2015ggf, Clough:2016ymm, Clough:2017efm, Aurrekoetxea:2019fhr, Joana:2020rxm} the full 3+1 equations of general relativity were solved numerically for the case of single scalar field inflation, and it was shown that in the large-potential scale case inflation is robust even against perturbations that reach the minimum of the potential. This includes models that are consistent with observations \cite{Planck:2018jri} such as Starobinsky inflation \cite{Starobinsky:1980te}. In the small-potential scale case, an analytic criterion was proposed for the maximum perturbation size that allows inflation to occur, and this was confirmed numerically \cite{Aurrekoetxea:2019fhr}. Similar results have been found for initial data that includes kinetic inhomogeneities \cite{Corman:2022alv, Elley:2024alx}, where black holes may form but some region of the spacetime remains inflationary. Some more recent work \cite{Garfinkle:2023vzf,Ijjas:2024oqn} has found conflicting results, calling for further investigations into the effects of the choice of initial data and its dependence on boundary conditions, see \cite{Joana:2024ltg} for a response to some aspects of these concerns and \cite{Kallosh:2025ijd} for a proposed explanation of the differences between \cite{Garfinkle:2023vzf,Ijjas:2024oqn} and the existing literature. A more detailed review of the current status of the application of NR to early universe cosmology is provided in \cite{Aurrekoetxea:2024mdy}.

In this work we continue studying the behaviour of large perturbations in the inflaton field by including the effects of higher-derivative terms in the gravitational action. The addition of these terms is motivated by the idea that classical General Relativity is the low-energy limit of some other UV-complete theory, and therefore there should be additional higher-derivative terms in the action that become relevant above some (as yet unknown) energy scale. Early in inflation these energy scales may be explored, and if these additional terms have important effects they may change our conclusions about the robustness of certain models \cite{Weinberg:2008hq}. In modern cosmology, many investigations into the phenomenology of modified gravity have been concerned with the longest observable length scales (i.e., late-universe Hubble scales), with a view to addressing late-time accelerated expansion or the dark matter problem. However, due to the strong constraints on modifications to gravity at shorter length scales like those of solar system tests, these theories often require the inclusion of screening mechanisms to turn off effects in certain environments \cite{Koyama:2015vza,Ferreira:2019xrr,Baker:2020apq}. On much higher curvature scales (length scales below 1 km), modified gravity effects are less constrained, and in the absence of sub-solar mass primordial black holes, compact objects will not be able to probe them. These considerations motivate us to investigate higher derivative effects in the early universe, where shorter length scales become dynamically relevant. For example, the Hubble scale during inflation can be of order $10^{-23}$m, and strong perturbations of the spacetime are therefore expected to exist on this scale at the start of inflation (or may be generated by non-linear effects during inflation, e.g. \cite{Caravano:2024tlp, Caravano:2024moy}), before they are smoothed out by the expansion.

As a starting point we consider what is arguably the simplest case - where a single scalar field is both the inflaton that provides the slow roll potential and the additional scalar degree of freedom of the metric. Whilst this is the simplest case, it restricts quite considerably the parameter space we can explore, as we will discuss further below. However, it still provides a useful illustration of the main effects of the additional terms, and allows us to highlight some of the challenges of ensuring a fair comparison when studying different initial conditions in modified gravity scenarios. In particular, we will be interested in whether higher derivative corrections can affect the behaviour of large perturbations (beyond effects that are captured in the near homogeneous and isotropic limit that have been well studied, see e.g. \cite{Kanti:1998jd,Nojiri:2005vv,Jiang:2013gza,Kanti:2015pda,Hikmawan:2015rze,vandeBruck:2015gjd,Nojiri:2017ncd,Sberna:2017xqv,Yi:2018gse,Odintsov:2018zhw,Chakraborty:2018scm,Fomin:2020hfh,Yogesh:2025wak}) while remaining in the weakly coupled regime, and whether non-linear dynamics can drive the system out of this regime. Broadly we conclude that (subject to some caveats) there is little change in the robustness of the model to perturbations when the additional higher derivative terms are included, even in the fully non-linear regime, and that only very finely tuned initial data can drive the model out of the weak coupling regime. This conclusion is not unrelated to the fact that an inflationary spacetime is close to de Sitter, and the latter is non-linearly stable \cite{Friedrich:1986qfi}. In this case, the exponential decay of the deviations from de Sitter is crucial to establish the non-linear stability of such spacetimes. 

In the future one could extend this to the case where two scalars separately provide the inflaton and the modification to gravity, and explore a wider class of couplings, which could give more interesting dynamics whilst remaining in the weak coupling regime. We will comment on this further in the discussion in Sec. \ref{sec:discuss}. 

This paper is organised as follows; in Sec. \ref{sec:model} the inflationary model will be described, including the modifications to general relativity and the form of the inflationary potential. The restrictions imposed by the homogeneous and isotropic limit for this model will then be discussed in Sec. \ref{sec:homogeneous_GB}. Restrictions imposed by weak coupling are discussed in Sec. \ref{sec:WCC}. In Sec. \ref{sec:initial_conditions} the initial data for our simulations will be described, along with a brief description of the methods used to solve the constraint equations in these theories. Our results, and some specific examples that demonstrate the key results, are then given in Sec. \ref{sec:results}. We conclude in Sec. \ref{sec:discuss}.

\section{Scalar-tensor model studied}\label{sec:model}

%We consider the most general parity-invariant four-derivative scalar-tensor effective field theory\footnote{This is a Horndeski theory, so the equations are second order in time and space, but contain terms up to fourth order in the derivative expansion, i.e. operators up to mass dimension $4$.}, up to field redefinitions \cite{Weinberg:2008hq}, which corresponds to the addition of a Gauss-Bonnet term coupled to a scalar field with a general potential $V$ and two kinetic terms. \hl{This theory is obtained by including in the action all generally covariant terms with up to four spacetime derivatives, removing total derivatives, and employing the equations of motion from the leading order terms.} Specifically, the action for this theory is

In this paper we focus on the effective field theory (EFT) for single field inflation as proposed by Weinberg \cite{Weinberg:2008hq} which we now summarise briefly. We consider a general low energy effective action containing all possible covariant interactions built out of a scalar field, the metric tensor and their derivatives. We assume that each term in the effective action is suppressed by a suitable negative power (determined by dimensional analysis) of a fundamental mass scale that characterises the UV physics. This way we can organise the possible interactions in a derivative expansion where operators are ranked by the number of derivatives they contain. At low energies (compared to the fundamental mass scale), the most important terms are those with the fewest derivatives, such as the scalar potential (containing zero derivatives), the Einstein-Hilbert term and the scalar kinetic term (containing two derivatives of the metric and the scalar field, respectively). Therefore, the omission of higher order terms in the derivative expansion is justified at low energies. Nevertheless, if we want to explore the possible effects that fundamental physics might have on inflationary dynamics we could retain operators up to higher than second order in the effective action. We shall refer to EFTs whose Lagrangian contains operators up to derivative order $n$ as `$n$-derivative theories' (even though some `$n$-derivative theories' with $n> 2$ may have second order equations of motion).

In the EFT of single field inflation the leading corrections to the Lagrangian of the minimally coupled Einstein-scalar-field theory are operators with four derivatives. The most general set of such operators is detailed in \cite{Weinberg:2008hq}. It is also shown in \cite{Weinberg:2008hq} that most of these operators are redundant in the sense that they are either total derivatives or they can be eliminated by using field redefinitions. The effective action simplifies even more if we additionally assume that our theory is invariant under spacetime parity transformations. The effective action we end up with is then
\begin{eqnarray}\label{eq:action}
S_{4\partial ST} = \int d^4x \sqrt{-g}\big[ R -V(\phi) + X + g_2(\phi)X^2 \nonumber \\  
+ ~ \lambda(\phi)\mathcal{L}^{\text{GB}}\big] ~,
\end{eqnarray}
where 
\begin{equation}
    X = - \frac{1}{2}(\nabla_\mu\phi)(\nabla^\mu \phi)~,
\end{equation}
\begin{equation}
    \mathcal{L}^{\text{GB}} = R^2-4R_{\mu\nu}R^{\mu\nu}+R_{\mu\nu\rho\sigma}R^{\mu\nu\rho\sigma} ~,
\end{equation}
and $g_2(\phi)$ and $\lambda(\phi)$ are smooth functions of the scalar field $\phi$.

The theory \eqref{eq:action} can thus be considered the most general parity-invariant four-derivative scalar-tensor effective field theory\footnote{To reiterate our terminology, we note that this theory is four derivative in the sense of the derivative expansion of the Lagrangian. This is not to be confused with the fact that \eqref{eq:action} is also a Horndeski theory, so the equations of motion of \eqref{eq:action} are second order in time and space derivatives.}, up to field redefinitions \cite{Weinberg:2008hq}.

To obtain a well-posed initial value problem for \eqref{eq:action} and to solve the equations of motion of this theory numerically, it is necessary to find an appropriate gauge choice and gauge-fixing procedure \cite{Kovacs:2020pns,Kovacs:2020pns,AresteSalo:2022hua,AresteSalo:2023mmd}. Stable numerical simulations further require the addition of suitable constraint-damping terms, see \cite{AresteSalo:2022hua,AresteSalo:2023mmd} for more details. The gauge-fixed equations of motion we solve may be written as follows:
\begin{align}
    &R^{\mu\nu}-\textstyle\frac{1}{2}Rg^{\mu\nu}-\hat P_\alpha^{~\beta\mu\nu}\nabla_\beta C^\alpha \\
    &+ \kappa_1 [n ^{(\mu}C^{\nu)}+\textstyle\frac{1}{2}\kappa_2 n^\alpha C_\alpha g^{\mu\nu}] = T^{\phi~\mu\nu} - 4\mathcal{H}^{\mu\nu}, \nonumber\\
    &\square \phi[1+2g_2(\phi)X]-V'(\phi)-3X^2g_2'(\phi)\\
    &-2g_2(\phi)(\nabla^\mu\phi)(\nabla^\nu\phi)\nabla_\mu\nabla_\nu\phi=-\lambda'(\phi)\mathcal{L}^{\text{GB}}\nonumber
\end{align}

where
\begin{align}
    &T^\phi_{\mu\nu} = \textstyle\frac{1}{2}\{(\nabla_\mu\phi)(\nabla_\nu\phi)(1+2g_2(\phi)X)\\
    &+g_{\mu\nu}[g_2(\phi)X^2+X-V(\phi)]\},\nonumber\\
    &\hat P_\alpha^{~\beta\mu\nu}=\delta_\alpha^{~(\mu}\hat g^{\nu)\beta}-\textstyle\frac{1}{2}\delta_\alpha^{~\beta}\hat g^{\mu\nu}, \\
    &H_{\mu\nu}=2R_{~(\mu}^{\rho}\mathcal{C}^{\mathstrut}_{\nu)\rho}-\mathcal{C}(R_{\mu\nu}-\textstyle\frac{1}{2}Rg_{\mu\nu})-\textstyle\frac{1}{2}R\mathcal{C}_{\mu\nu} \\
    & +  \mathcal{C}^{\alpha\beta}(R_{\mu\alpha\nu\beta}-g_{\mu\nu}R_{\alpha\beta}), \nonumber\\
    &\mathcal{C}_{\mu\nu}\equiv\lambda'(\phi)\nabla_\mu\nabla_\nu\phi+\lambda''(\phi)(\nabla_\mu\phi)(\nabla_\nu\phi), \\
    &\mathcal{C}^\mu=H^\mu+\tilde g^{\rho\sigma}\Gamma^\mu_{\rho\sigma}.
\end{align}
$H^\mu$ are
the source functions that parametrize the underlying coordinate freedom of the theory, and $\mathcal{C}\equiv g^{\mu\nu}\mathcal{C}_{\mu\nu}$. The auxiliary metrics are given by
\begin{align}
    \tilde g^{\mu\nu}=g^{\mu\nu}-a(x)n^\mu n^\nu,~~~
    \hat g^{\mu\nu}=g^{\mu\nu}-b(x)n^\mu n^\nu,
\end{align}
where $n^\mu$ is the unit normal to surfaces of constant time, $a(x)$ and $b(x)$ are chosen such that $0<a(x)<b(x)$ or $0<b(x)<a(x)$, and $\kappa_1$ and $\kappa_2$ are damping coefficients.

There are many possible choices for the inflationary potential, since the only requirements are a sufficiently long plateau and a final minimum (set here at $\phi = 0$). In this work we choose the family of $\alpha$-attractor models \cite{Kallosh:2013hoa, Kallosh:2013yoa} (of which the Starobinsky model is a specific case), with the potential in these models given by,
\begin{equation}\label{eq:Starobinsky_pot}
V(\phi) = \Lambda^4(1-e^{\phi/\mu})^2 ~.
\end{equation}
Here $\Lambda$ sets the energy-scale of inflation, and can be chosen to satisfy observational constraints from the primordial density power spectrum. Since the inflection point of the potential occurs at $\phi = -\ln(2)\mu$, the parameter $\mu$ then dictates whether the potential scale is small or large. This model was shown in \cite{Aurrekoetxea:2019fhr} to be robust against large inhomogeneities in the large potential scale case, and to become less robust as that scale decreases. In this work we take the intermediate value $\mu = 0.1 M_{pl}$. 

For the other model parameters we set $g_2(\phi)=g_2$ and $\lambda(\phi)=\lambda \phi$, where $\lambda$ and $g_2$ are dimensionful coupling constants. We will mostly focus on the case where $g_2=0$, but will return to the more general case of a constant, non-zero value in section \ref{sec:g2}. 

\begin{figure}[!ht]
    \begin{center}
    \includegraphics[width=0.6\textwidth]{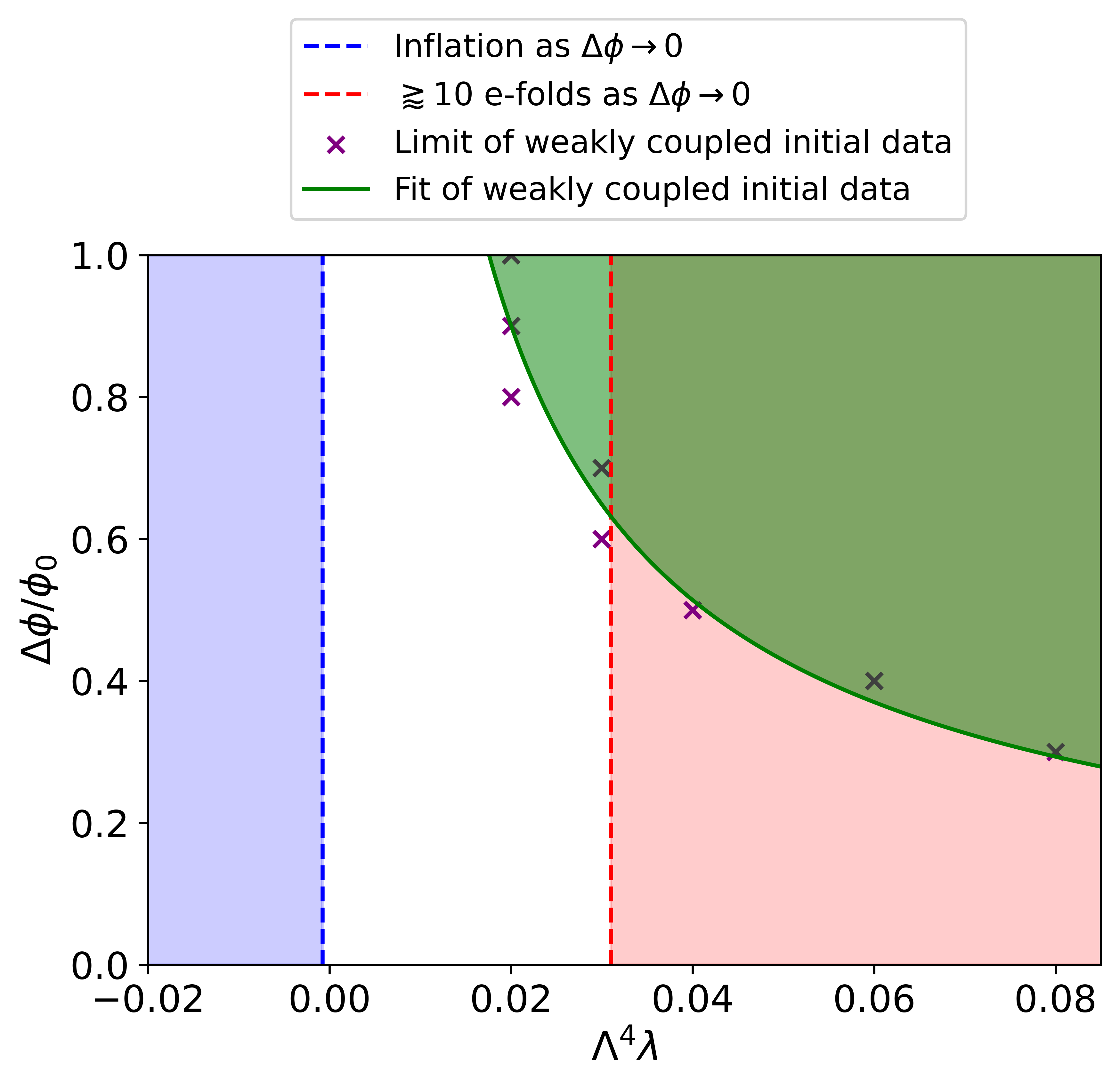}
     \caption{Restrictions on combinations of the relative initial perturbation size $\Delta\phi/\phi_0$ and the Gauss-Bonnet coupling $\lambda$ for $N=2$ modes per (FLRW) Hubble length (see Section \ref{sec:initial_conditions} for definitions), with exclusion zones set by requiring the success of inflation as $\Delta\phi \rightarrow 0$ and weak coupling in the initial data. This leaves only the white area for study in this work.}
     \label{fig:limits}
     \end{center}
\end{figure}

While many different combinations of potentials and coupling functions may provide the required 60 or more e-folds of inflation, there are some constraints that limit the space we can investigate for the purposes of this study. Firstly, we would like that inflation succeeds as the scalar perturbations are taken to zero, which restricts the coupling to a specific range. We also require that the higher derivative terms are sufficiently small so that the theory is initially in the regime in which the EFT is valid, where it should admit a well-posed initial value formulation\footnote{This condition is also required to hold in the recently proposed regularisation scheme of \cite{Figueras:2024bba}, an alternative approach which could also be applied to the class of theories studied here.}. In the following sections we consider the restrictions that these constraints impose, which are illustrated in Fig. \ref{fig:limits}.

It is not the goal of our work to provide specific model predictions, but to characterise the general behaviours that occur if one considers the higher order corrections in a general scalar-tensor theory in which the scalar is the inflaton. As above, our main questions are whether, given the above restrictions, the addition of the higher derivative terms tend to make the model more or less robust, and whether there are any non-linear dynamics that may drive the field out of the EFT regime.
Our work can be generalised to other choices for the couplings or inflationary potential, but we expect that these behaviours remain qualitatively similar in other regimes, with just the specific values changing for stability limits, max/min number of e-folds of inflation, etc. 

\section{Restrictions from the homogeneous limit}\label{sec:homogeneous_GB}

\begin{figure}[!ht]
    \begin{center}
    \includegraphics[width=0.6\textwidth]{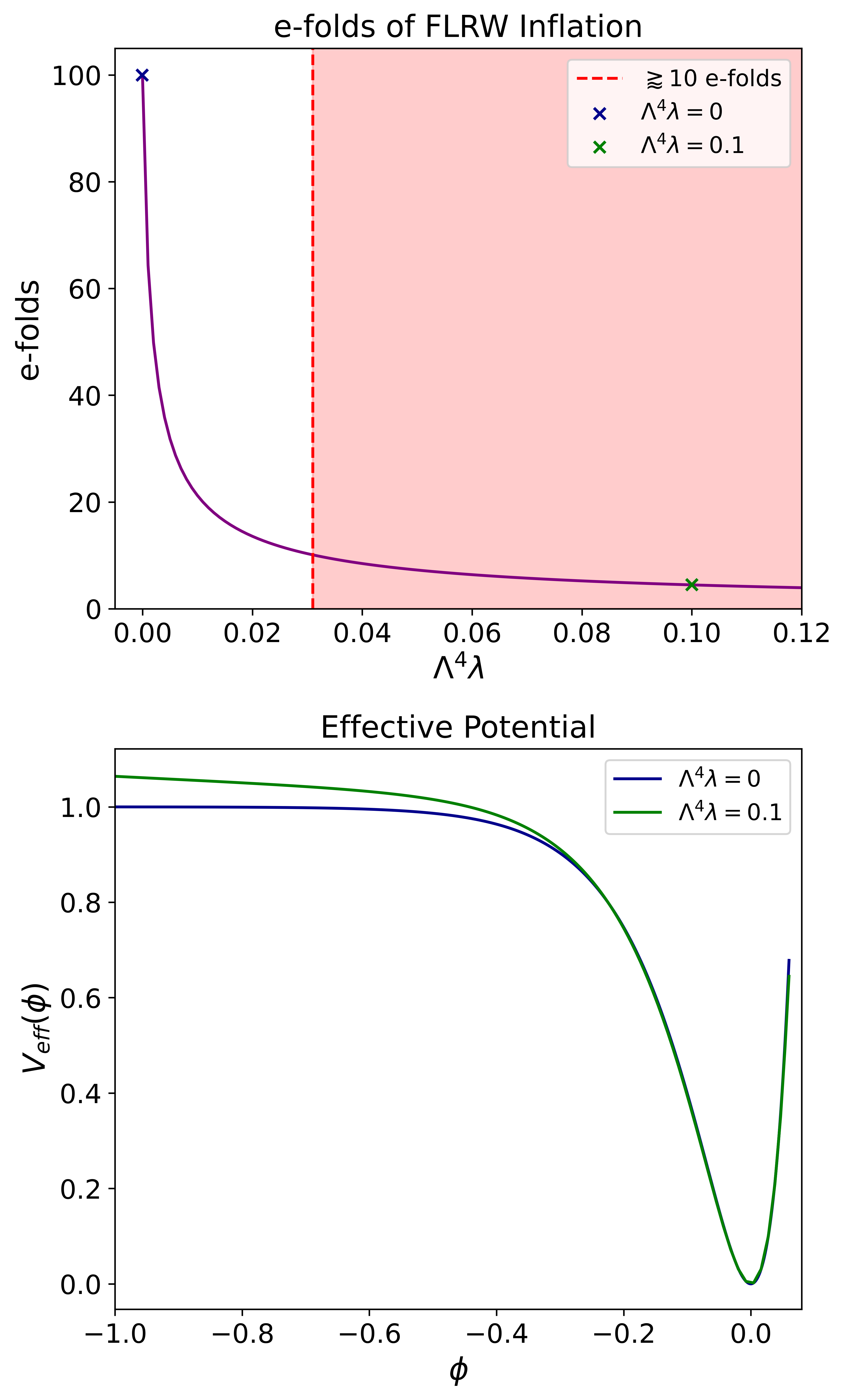}
     \caption{Top panel: The total number of e-folds of inflation against the coupling $\lambda$. We see that the number of e-folds is very sensitive to changes in the value of $\lambda$, since even a small change increases the slope of the effective potential.
     Bottom panel: Plots of the effective potential, $V_\text{eff}$, for two different values of the coupling, $\lambda$. We see that the effect of increasing the coupling is to increase the slope of the effective potential, thus potentially disrupting the slow roll condition.
     }
     \label{fig:efolds_Veff}
     \end{center}
\end{figure}

In the homogeneous and isotropic case, imposing spatial flatness, the spacetime can be described by the Friedmann-Lemaitre-Robertson-Walker (FLRW) metric,
\begin{eqnarray}
    ds^2 = -dt^2 + a(t)^2 (dx^2 + dy^2 + dz^2)
\end{eqnarray}
The full field equations (given in this formulation by equations (2)-(4) of \cite{AresteSalo:2023mmd}) then reduce to coupled ODEs
\begin{align}
    &6H^2 - \frac{1}{2}\dot{\phi}^2-V(\phi) + 24\lambda'(\phi)\dot{\phi}H^3 = 0 ~, \label{eq:FLRW1} \\
    &\ddot\phi + 3H\dot\phi + V'(\phi)-24\lambda'(\phi)\big(\dot{H} H^2 + H^4\big) = 0 ~, \label{eq:FLRW2}
\end{align}
where the prime represents differentiation with respect to $\phi$. Here we neglect the four-derivative scalar term proportional to $g_2(\phi)$, which would contribute terms of order $\dot\phi^2$ or higher. Assuming the slow-roll conditions are satisfied, these terms are suppressed during inflation, and should therefore not play a significant role. The same argument can be made for other components from the Gauss-Bonnet contribution in Eq. \eqref{eq:FLRW1}, leaving only the term proportional to $\lambda'(\phi)$ in Eq. \eqref{eq:FLRW2} in addition to the usual GR terms. The modification can then be described by an effective potential for the scalar field during inflation of the form
\begin{equation}
    V_{\text{eff}}(\phi) = V(\phi)-24\lambda(\phi)(\dot H H^2 + H^4).
\end{equation}
The effect of this change on the potential shape is illustrated in the bottom panel of Figure \ref{fig:efolds_Veff}, and the top panel shows the dependence of $N_\text{final}$, the total number of e-folds of inflation, on $\lambda$, the coupling parameter. We see that the scalar and spacetime behaviour is very sensitive to changes in the value of $\lambda$, since even a small change increases the slope of the effective potential. 

With a linear coupling function and our chosen potential, the requirement that our models provide successful inflation as the perturbation size is taken to zero (i.e., in the homogeneous case) introduces the following condition on the coupling parameter $\lambda$,
\begin{equation}\label{eq:lambda_constraint}
    \lambda > \frac{-3}{\Lambda^4\mu}\frac{e^\frac{\phi_0}{\mu}}{(1-e^{\frac{\phi_0}{\mu}})^3}
\end{equation}
where $\phi_0$ is the initial value of the scalar field. If this condition is not met, the scalar field will initially accelerate away from the minimum of the potential and inflation will not end. Equally if the value of $\lambda$ is too large, then slow roll will be disrupted and fewer e-foldings of inflation will occur before we reach the potential minimum. We choose to impose a cut off at the value giving only 10 e-folds (a somewhat arbitrary choice). These restrictions give the vertical limit lines shown in Fig. \ref{fig:limits}, which define the blue and red exclusion regions respectively. Stability analyses of these inflationary FLRW solutions are given in \cite{Guo:2006ct,Leith:2007bu,Sberna:2017nzp,Odintsov:2023lbb}, where it is shown that the self-interaction (potential) term is sufficient to cure the instability of tensor perturbations that can occur in EsGB theory in expanding spacetimes.

Analytic predictions for the behaviour of the homogeneous spacetime should also hold with small enough perturbations and Gauss-Bonnet couplings, so we use them to validate our numerical implementation, in particular the unit conventions, as discussed in Appendix \ref{sec:appendix}.

\section{Restrictions from the weak coupling condition}\label{sec:WCC}

Given the effective field theory motivation for the model studied here (that is, it is assumed that 5th order and higher derivative terms that could be included in the action Eq. \eqref{eq:action} are suppressed and can be neglected), then we need the correction terms to be subleading to the two-derivative theory to allow the description to remain valid. From a more pragmatic perspective, starting within the weakly coupled regime ensures that, at least initially, in some local neighbourhood of the initial hypersurface, the evolution equations remain strongly hyperbolic and can be numerically evolved \cite{Kovacs:2020ywu, Kovacs:2020pns}. Note that imposing this restriction on the initial data does not guarantee that it remains satisfied at all times - dynamical evolution may drive the fields out of the weakly coupled regime, as has been observed in studies of black holes in the same formalism
\cite{Doneva:2024ntw,Doneva:2023oww,Thaalba:2024crk,Thaalba:2024htc,Thaalba:2023fmq,Franchini:2022ukz,East:2020hgw,East:2021bqk,Corman:2024vlk, Corman:2022xqg,Corman:2024cdr,Ripley:2019aqj,Ripley:2019aqj,AresteSalo:2022hua,AresteSalo:2023mmd,R:2022hlf}.

The weak coupling condition (WCC) quantifies to what extent the spacetime is in the EFT regime. It requires that the length scale associated with the scalar-Gauss-Bonnet coupling is small compared to the other relevant length scales in the system. In practice, we require that\footnote{Note that the definition of the characteristic length scale $L$ and the WCC given here is not covariant. One could alternatively use a covariant definition based on the relative size of the four-derivative terms compared to the two-derivative terms in the Lagrangian. However, this latter definition would not take into account possible differences between different directional derivatives, and hence, the definition \eqref{eq:char_L} is more conservative. In our simulations these two definitions of $L$ show good agreement.},
\begin{eqnarray}\label{eq:WCC}
L^{-1}\sqrt{|\lambda'(\phi)|} \ll 1,
\end{eqnarray}
where,
\begin{eqnarray}\label{eq:char_L}
    L^{-1}=\max\{|R_{\mu\nu\rho\sigma}|^\frac{1}{2}, |\nabla_\mu\phi|,
    |\nabla_\mu\nabla_\nu\phi|^\frac{1}{2}, |V(\phi)|^\frac{1}{2}\}.
\end{eqnarray}

Weak coupling is also closely related to the question of whether we can solve the constraint equations for the initial data \cite{Kovacs:2021lgk}.
In the CTTK (Conformal Transverse-Traceless-K) formalism we use \cite{Aurrekoetxea:2022mpw}, the method of choosing the extrinsic curvature $K$ to solve the Hamiltonian constraint requires that the effective energy density must be non-negative everywhere, which is only guaranteed if the contributions from the usual GR scalar field energy density are greater than those introduced by the Gauss Bonnet coupling.
%$|\rho_{SF}| \geq |\rho_{GB}|$
Whilst using different choices for some of the variables (e.g. solving for a non constant conformal factor) can somewhat alleviate this restriction, we consistently find that this marks the point at which unique solutions can no longer be found and/or the evolution rapidly breaks down.
We illustrate this restriction on the initial data in Figure \ref{fig:limits}. For a fixed initial value of $\phi_0$ that gives 100 e-folds of inflation in the absence of perturbations, the requirement of weak coupling in the initial data places a restriction on the coupling $\lambda$ that depends on the perturbation size $\Delta\phi$ (as defined in the following section). As expected, $\lambda$ is unbounded from above as the perturbation size goes to zero, and the maximum allowed value of $\lambda$ is inversely related to the perturbation size.

We note that the above measure has been employed in previous works studying black hole spacetimes, and is generically successful at predicting where the numerical scheme will break down (see e.g. \cite{R:2022hlf,Doneva:2023oww,AresteSalo:2022hua,AresteSalo:2023mmd,Ripley:2019irj}). However, in the special case of (nearly) homogeneous cosmological spacetimes the condition for the breakdown of solutions appears to be weaker. Here the LHS of Eq. \eqref{eq:WCC} may be $O(10^{-1})$ while the perturbations die away, but it increases to $O(1)$ values later in the evolution as the $\dot{\phi}$ term becomes large at the end of inflation. At this point, the evolution can be described by the ODEs in equations \eqref{eq:FLRW1} and \eqref{eq:FLRW2}, whose principal (highest derivative) terms are well-behaved\footnote{Note, in particular, that \eqref{eq:FLRW1} is a quadratic algebraic equation for $\dot\phi$ whose discriminant is positive under the same conditions as in the minimally coupled theory, i.e. regardless of the size of $\lambda'(\phi)$. On the other hand, global well-posedness for \eqref{eq:FLRW1}-\eqref{eq:FLRW2} still depends on the structure of the lower order terms. A careful mathematical analysis of this problem is beyond the scope of this paper.}.
Nevertheless, one should keep in mind that \eqref{eq:WCC} is relevant for evaluating the validity of the theory, especially since we want to focus on the inhomogeneous part of the evolution.

\section{Initial data}
\label{sec:initial_conditions}

There are 16 variables that must be set in the initial conditions and four constraint equations, resulting in a highly under-determined system. As a result, a well-motivated separation must be made between variables that are chosen, and others that are solved for (using the constraint equations). 

In this work a modified version of the GRFolres solver \cite{Aurrekoetxea:2025kmm} is used to solve the constraint equations, using the CTTK method \cite{Aurrekoetxea:2022mpw}. The adaptation of this method to the modified gravity case is described in detail in \cite{Brady:2023dgu}, building on the analysis of \cite{Kovacs:2021lgk}. Here we simply highlight the choices relevant to this work.

In the CTTK method, we choose the scalar field configuration $\phi$ and its conjugate momentum $\Pi$, the conformal three-metric $\bar\gamma_{ij}$, and the transverse-traceless part of the conformal extrinsic curvature, $\bar A^{TT}_{ij}$. We then use the Hamiltonian constraint to solve for the mean curvature $K$, and use the momentum constraints to solve for the longitudinal part of the extrinsic curvature, $\bar A^{L}_{ij}$. The conformal factor $\psi$ can be set in two ways --- as a free choice in the Hamiltonian constraint (in which case $\psi=1$ is usually chosen), or as another dependent variable used to account for some of the terms in the same constraint. This method is adapted for EsGB gravity by considering all the additional terms in the Hamiltonian constraint as contributions to an effective energy density, combining them with the energy density of the scalar field. Likewise in the momentum constraints, the additional terms coming from the higher derivative modifications are considered as contributions to an effective momentum density. The validity of this approach was confirmed in \cite{Brady:2023dgu}, where convergence tests showed that the method solves the constraints to the expected order.

In the initial data used for this work, the metric is assumed to be conformally flat (i.e., $\bar\gamma_{ij}=\delta_{ij}$), the transverse-traceless part of the conformal extrinsic curvature $\bar A_{ij}$ is set to zero (roughly speaking, this means that there is no initial gravitational wave content). These choices are somewhat arbitrary but we do not expect them to materially affect our conclusions - conformal flatness is imposed on average by the periodicity of the domain, but the fact that it applies everywhere is a limitation of our method of solving the initial constraints. Similarly a zero contribution from the TT part of the extrinsic curvature is a simplification, but in previous works they have been shown not to strongly affect the scalar dynamics that determine the end of inflation \cite{Clough:2017efm}. The lapse $\alpha$ is set to one, and the shift $\beta_i$ is zero, but they will evolve away from these values since a dynamical gauge (1+log slicing) is used. For sufficiently small gradients in the field, the Gauss-Bonnet effective energy density contribution $\rho_{GB}$ should be subdominant to the scalar field density $\rho_{SF}$ everywhere. Then the Hamiltonian constraint can be solved analytically\footnote{One could also use the ``CTTK Hybrid'' method to solve for a non-constant $\psi$, which we find allows for the construction of initial data with slightly larger couplings. However, in the majority of these cases the solutions quickly break down once evolved, and so in practise end up being restricted in the same way as the simpler method we use with $\psi$ constant. Reducing $\psi$ can make the gradient energy smaller in some regions, so one is effectively just pushing the problem of not being weakly coupled into a different combination of variables.}, with both energy density contributions accounted for by setting $K^2 = 16\pi\rho$. 

The configurations of the scalar field and its momentum are heavily constrained by the integrability conditions. The momentum constraints are elliptic equations, and the sources must therefore integrate to zero when periodic boundary conditions are used. In the GR case, this means that there must be no net momentum flux over the periodic domain. In the case with EsGB modifications, the condition is much less easy to impose. Imposing zero scalar field momentum ($\Pi=0$) and a periodic scalar field configuration satisfies the conditions, but more work is needed to solve the constraints with more general configurations for $\Pi$ (and is likely to require removing the assumption of conformal flatness). The initial average value of the scalar field, $\phi_0$, is set to be the value that would give 100 e-folds in the homogeneous case with $\lambda=0$. This is chosen as an order-of-magnitude estimate of the required number of e-folds during inflation, and our results are not sensitive to this value. Using equations \eqref{eq:FLRW1} and \eqref{eq:FLRW2} above, this corresponds to an initial value of $\phi_0 = -1.06 M_{\textnormal{Pl}}$.

Scalar perturbations are then included by setting the initial scalar field profile to,
\begin{eqnarray}\label{eq:perturbations_form}
    \phi = \phi_0 + \frac{\Delta\phi}{3}\Big(\cos(kx)+\cos(ky)+\cos(kz)\Big) ~,
\end{eqnarray}
where $\Delta\phi$ is the maximum perturbation amplitude, and $k=\frac{2\pi N}{L}$ is the wavenumber. $N$ is an integer and $L$ is given by
\begin{eqnarray}
    L=\frac{1}{H_0}=\sqrt{\frac{6}{V(\phi_0)}} ~,
\end{eqnarray}
the initial Hubble radius when $\Delta\phi\rightarrow0$ and $\Pi=0$. This simple profile nevertheless provides a good toy model for horizon scale perturbations, which have been previously shown to be the ones that most affect the robustness of inflation \cite{Clough:2015sqa, East:2015ggf, Corman:2022alv}.

\section{Results}\label{sec:results}

The initial conditions described in Section \ref{sec:initial_conditions} can be evolved according to the equations of motion derived from the action, Eq. \eqref{eq:action}. This is done using the GRFolres numerical relativity code \cite{AresteSalo:2023hcp}, with the modified CCZ4 method described in \cite{AresteSalo:2022hua, AresteSalo:2023mmd}, which adapts the modified GHC formalism of \cite{Kovacs:2020ywu, Kovacs:2020pns} to the puncture gauge. Further details of the implementation are given in the Appendix \ref{sec:appendix}.

In the following sections we consider several cases of interest, and summarise the key results. 
In the first section, we fix the initial average value of the scalar field and vary the perturbation size and coupling strength. We then permit the initial average value of the field to vary, which allows us to restore sufficient inflation for even very large perturbations by starting the field further up the inflationary plateau. We then consider whether we can dynamically drive the field out of the weakly coupled regime of the EFT during the period over which it homogenises - finding that, whilst possible, this requires significant fine tuning. Finally we highlight the role of the four-derivative scalar term $g_2$, which is the term that most affects the initial homogenisation of large perturbations.

\subsection{Fixing the average field value $\phi_0$}\label{sec:large_perturbations}

\begin{figure}[!ht]    
    \begin{center}
    \includegraphics[width=0.6\textwidth]{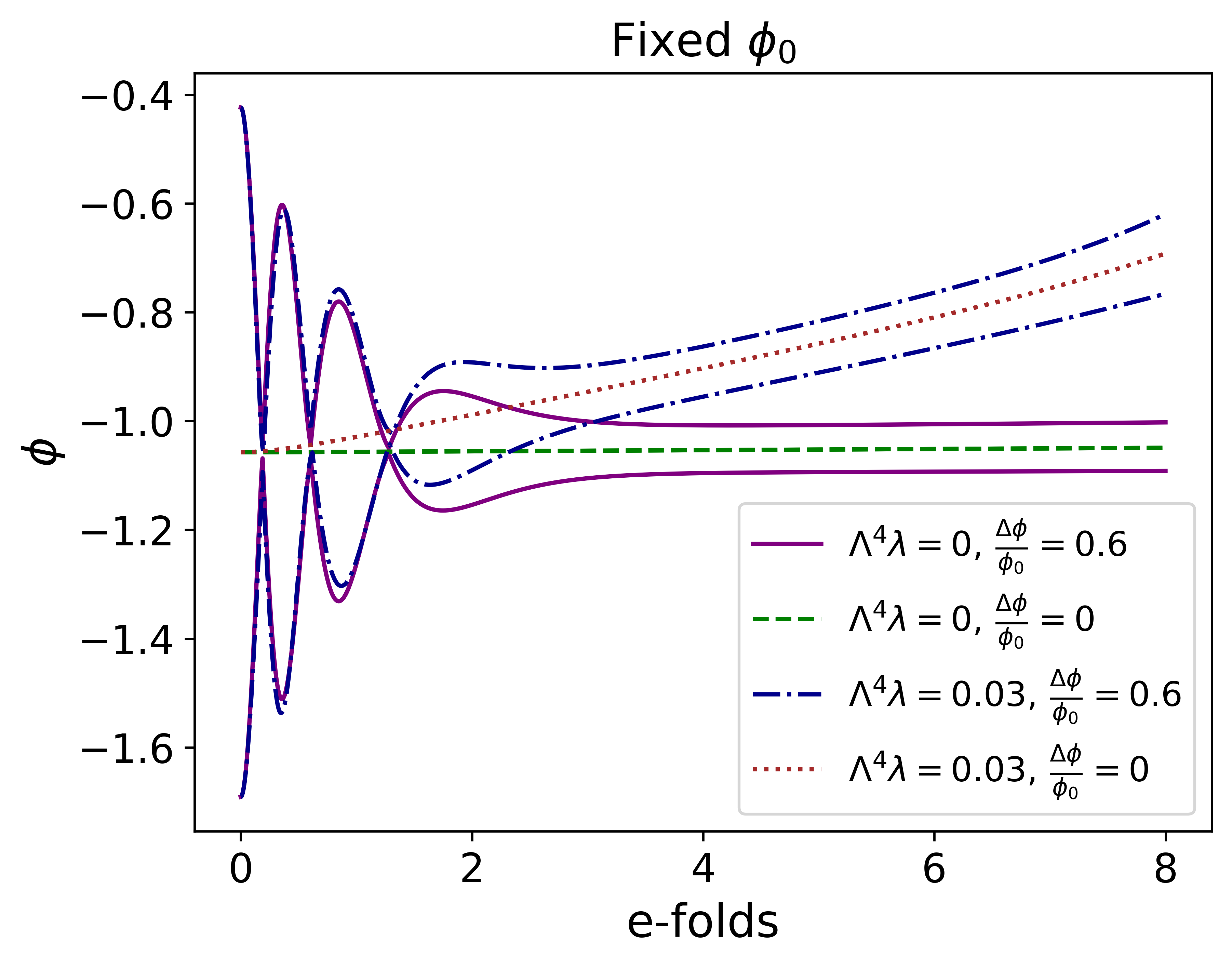}
     \caption{Plots of the maximum and minimum values of $\phi$ against N-folds of expansion, for both zero and non-zero coupling $\lambda$, and zero and non-zero perturbation size $\Delta\phi$. The initial average value $\phi_0$ is fixed to provide 100 e-folds of inflation with $\lambda=0$, and there are $N=2$ modes per Hubble radius. We see that the inclusion of the Gauss-Bonnet coupling significantly affects the homogeneous behaviour, but does not affect the decay of large perturbations.}
     \label{fig:fixedphi0}
     \end{center}
\end{figure}

We first choose a fixed average field value $\phi_0$ that gives 100 e-folds in the homogeneous GR case. Within the bounds shown in Fig. \ref{fig:limits} and discussed in Sec. \ref{sec:model}, the largest allowed coupling is then $\Lambda^4\lambda=0.03$, with relative perturbation sizes up to $\Delta\phi/\phi_0 \lessapprox 0.6$. With these restrictions, the deviations from the $\lambda=0$ case are dominated by the zeroth-order behaviour, as shown in Fig. \ref{fig:fixedphi0}.

In both the GR case (consistently with \cite{Aurrekoetxea:2019fhr}) and the case with non-zero $\lambda$, the scalar field perturbations initially decay. Inflation then continues with a small spread in $\phi$, which will eventually cause a small variation in the number of e-folds of inflation in different regions. The metric likewise shows short-lived perturbations in response to the inhomogeneities, but these decay, with the metric eventually becoming locally FLRW everywhere, with some variation left on super-horizon scales.
As discussed above, the condition of weak coupling is applied here in two ways. Firstly, in the initial data, the scalar field energy density $\rho_\text{SF}$ must dominate over the effective EsGB density contribution $\rho_\text{GB}$ everywhere. Secondly, during the evolution, the condition in Eq. \eqref{eq:WCC} is monitored. We find that cases which are very close to the limit for the former typically leave the weakly coupled regime early in the evolution, and the numerical scheme breaks down.

In summary, while the weak coupling conditions are obeyed, even very large horizon-scale modes are not significantly affected by the addition of the higher derivative terms. The overall effects are well described by the effective potential of the homogeneous case, after the initial dynamical evolution of the perturbations has died away. The perturbations of the field above the homogeneous value behave in a similar way to the GR case, with a similar decay rate of their amplitude before the mode freezes out.

\subsection{Varying the average field value $\phi_0$}

In principle, when varying the coupling $\lambda$, and hence the FLRW behaviour of the inflationary model, we are also free to adjust $\phi_0$ to ensure that we still obtain 100 e-folds of inflation in the homogeneous case with the effects of $\lambda$ included. That is, we can push the inflaton further up the plateau to compensate for the additional tilt imposed by the coupling. In this case, the upper limit set on $\lambda$ in Fig. \ref{fig:limits} by the requirement of sufficient e-folds (in the homogeneous limit) is avoided (but the lower one remains). However, the requirement of weakly coupled initial data still constrains the absolute size of $\Delta\phi$, leading to greater restrictions on $\Delta\phi/\phi_0$\footnote{The initial Hubble scale is nearly unaltered (due to the flatness of the potential in this region), so the same relative perturbation leads to much higher gradients and therefore larger modified gravity contributions in the WCC.}. Thus we end up with a similarly restricted plot, just with different specific values. 

\begin{figure}[!ht]
    \begin{center}
    \includegraphics[width=0.6\textwidth]{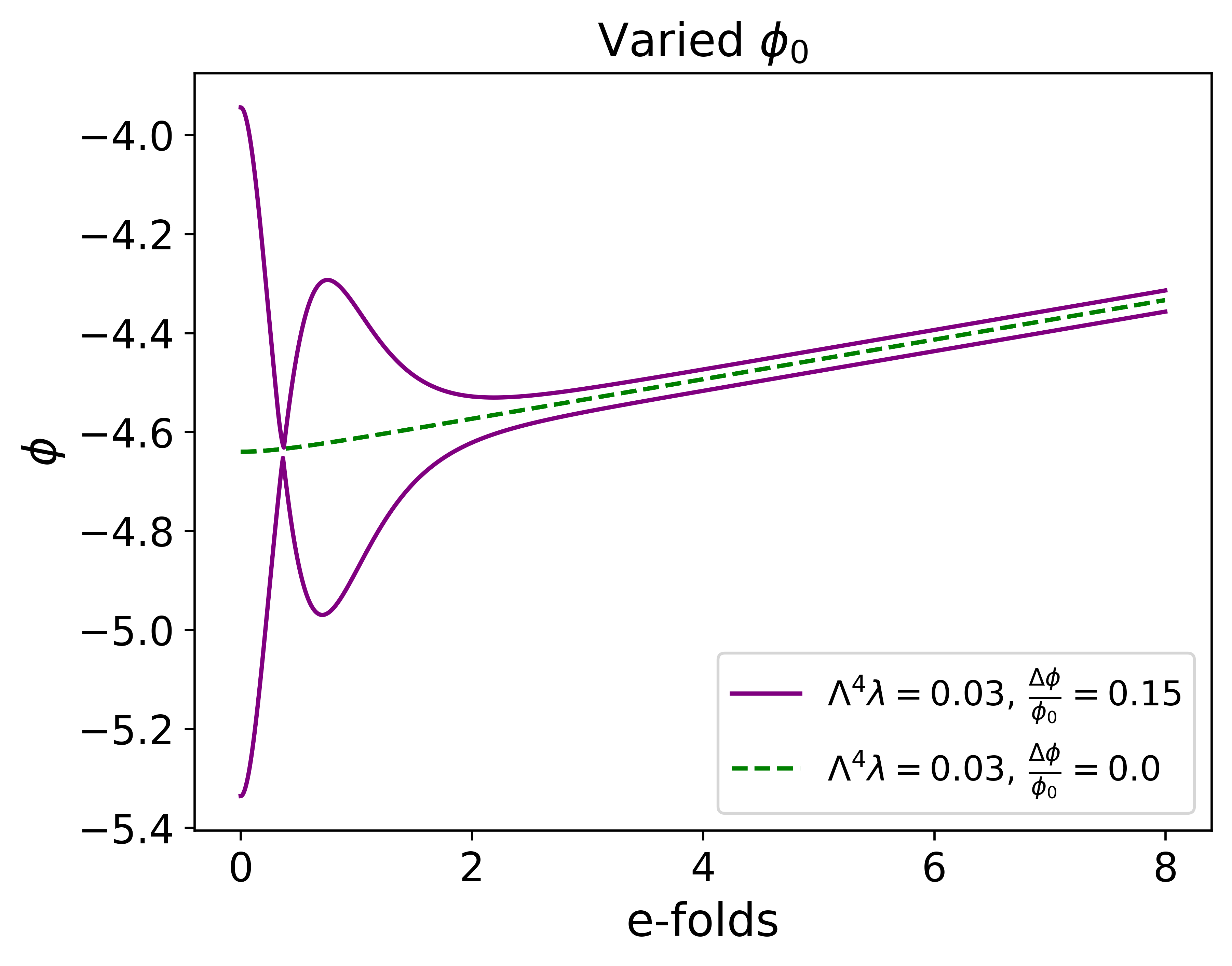}
     \caption{Plots of the maximum and minimum values of $\phi$ against N-folds of expansion, for fixed coupling $\lambda$, and zero and non-zero perturbation size $\Delta\phi$. The initial average value $\phi_0$ is varied to provide 100 e-folds of inflation with $\Lambda^4\lambda=0.03$, and there is $N=1$ mode per Hubble radius. The case with $\Lambda^4\lambda=0$ behaves similarly to the GR case shown in Figure \ref{fig:fixedphi0}. We see that the inclusion of the Gauss-Bonnet coupling does not significantly alter the behaviour of perturbations, which decay within the first few e-folds.}
     \label{fig:fixedlambda1}
     \end{center}
\end{figure}

The result is then consistent with the case where $\phi_0$ was fixed with respect to $\lambda$, such that the overall effect of the modified gravity terms is still well captured by the effective potential, after the initial dynamical evolution of the perturbations has died away. An example of this is shown in Fig. \ref{fig:fixedlambda1}.

\subsection{Can dynamical evolution break the WCC?}

In both of the above cases, the effect of the higher-derivative terms on the perturbations is minimal. An explanation for this is that in both cases the behaviour must initially be dominated by GR effects, which cause a decrease in the gradients, and therefore a decrease in the contribution of the modified gravity terms that depend on spatial gradients. Finding a scenario where the additional terms make an important contribution while still remaining initially weakly coupled would therefore require a high degree of fine-tuning.

\begin{figure}[!ht]
    \begin{center}
    \includegraphics[width=0.6\textwidth]{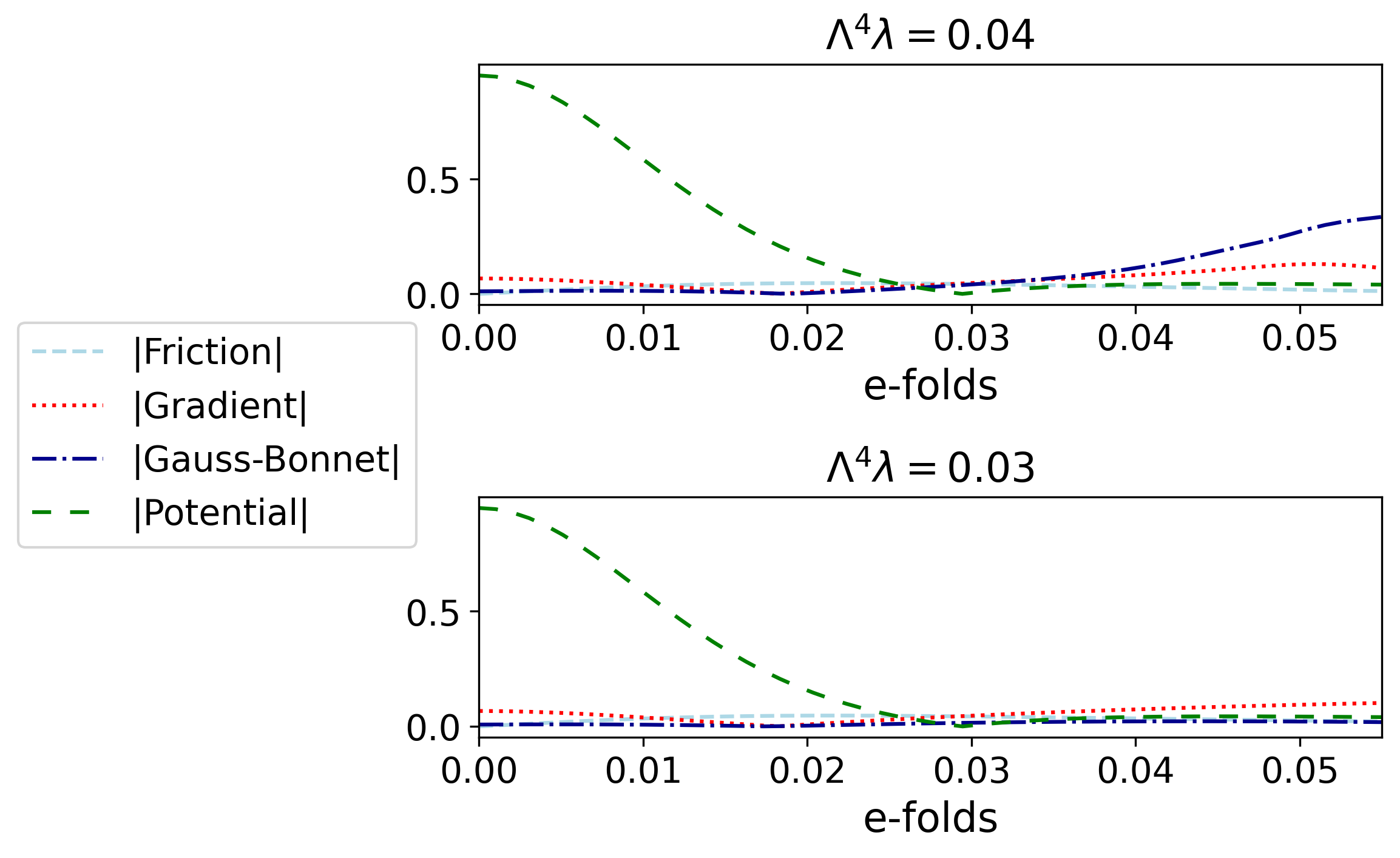}
     \caption{Plots of the different contributions to $\frac{\partial^2\phi}{\partial t^2}$, relative to its initial magnitude, against e-folds for two different values of the coupling $\lambda$. The explicit form of the contributions is listed in the appendix. In the first panel we are able to create a scenario where the Gauss-Bonnet contribution (blue dot-dash-line) dynamically grows such that it dominates over the usual GR terms, despite starting as a subdominant contribution. However, a small change in the coupling results in a different evolution where it remains sub dominant, as shown in the second panel. We conclude that the setup would need a high degree of fine-tuning for initially weakly coupled data to be driven out of that regime.}
     \label{fig:fine_tuned_plots}
     \end{center}
\end{figure}

We can achieve this by engineering a scenario near the threshold of weak coupling, where the gradients initially grow. Figure \ref{fig:fine_tuned_plots} shows one example of this. In this case, the scalar field initially explores beyond the minimum of the potential, leading to the temporary growth of gradients in that region when the maximum value of the scalar field accelerates to negative values (as opposed to the cases described above, where gradients initially decay). This can cause a growth in the Gauss-Bonnet contribution to $\partial^2_t\phi$, as shown in the lower panel of Figure \ref{fig:fine_tuned_plots}. As expected, the LHS of \eqref{eq:WCC} is order one here, and the numerical evolution quickly breaks down, even though the gradients begin to decay. With slightly lower values of the coupling $\Lambda^4\lambda$, as shown in the upper panel of Figure \ref{fig:fine_tuned_plots}, the evolution persists but the Gauss-Bonnet contribution to the dynamics is minimal. This shows that the coupling parameter would have to be very fine-tuned to remain in the weakly coupled regime while making a significant contribution to the dynamics, even in regions where gradients initially grow.
In summary, we conclude that generically the existence of the additional higher derivative terms does not dynamically drive the evolution out of the regime of the EFT, except in very finely tuned cases.

\subsection{Influence of the fourth order kinetic term $g_2$}\label{sec:g2}

In previous numerical simulations of these theories, which have investigated the dynamics of black hole systems, the four-derivative scalar term (proportional to $g_2(\phi)$ in equation \eqref{eq:action}) is usually found to have minimal effects \cite{AresteSalo:2022hua, AresteSalo:2023mmd}, since gradients remain small and subdominant.

However, in this scenario, the scalar field can have highly non-trivial dynamics before the perturbations die away, and this term may therefore have noticeable effects on the final state, well before it results in a violation of the weak coupling condition for this term (given by Eq. \eqref{eq:WCC} with $|\lambda'(\phi)|$ replaced by $|g_2(\phi)|$).

\begin{figure}[!ht]
    \begin{center}
    \includegraphics[width=0.6\textwidth]{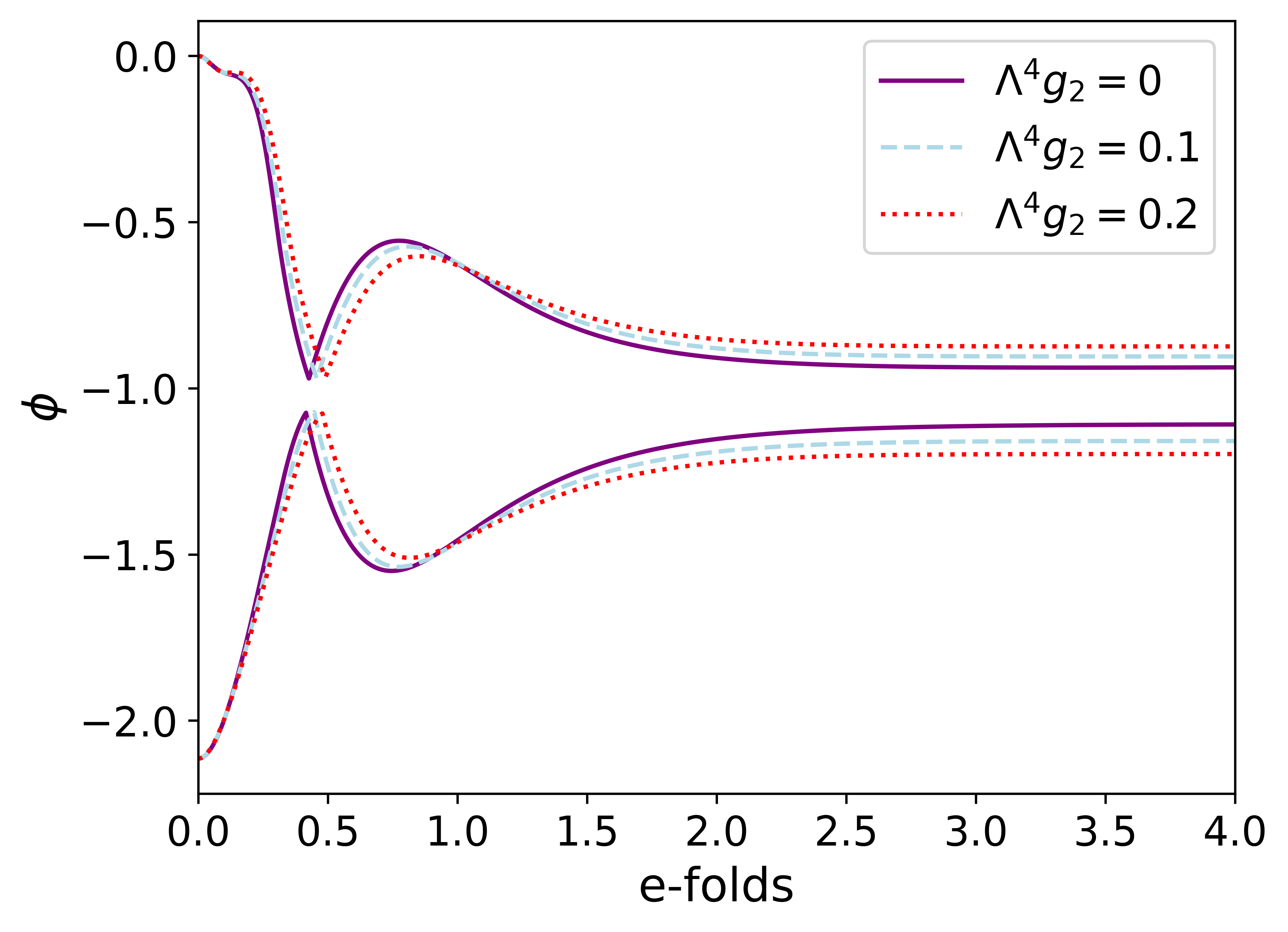}
     \caption{Plots of the maximum and minimum values of $\phi$ against N-folds of expansion, for three values of the coupling $g_2$ and a fixed perturbation size reaching the minimum of the potential, $\Delta\phi=\phi_0$. We see that the $g_2$ coupling does not affect the homogeneous behaviour, but does change the final amplitude of large perturbations.}
     \label{fig:g2_plots}
     \end{center}
\end{figure}

An example of these effects is shown in Fig. \ref{fig:g2_plots}. The function $g_2(\phi)$ is taken to be a constant, $g_2$, and this is varied up to a value where weak coupling is violated. The initial profile of the scalar field is the same as in section \ref{sec:large_perturbations}, with large perturbations reaching the minimum of the potential and an average value of $\phi$ that gives 100 e-folds of inflation in the GR case.

The average behaviour is largely unaffected by the inclusion of this term, as expected. However, the final amplitude of the scalar field perturbations is noticeably affected, with larger values of $g_2$ leaving larger final amplitudes in the mode after it freezes out. Therefore we conclude that this term will potentially make models less robust to large initial perturbations. 
Different signs of $g_2$ affect the scalar dynamics in different ways, in particular, the nature of the change of character of the equations (Tricomi vs Keldysh-type), as studied in \cite{Ripley:2019irj,Bernard:2019fjb,Bezares:2021dma,Figueras:2020dzx,Figueras:2021abd}. We found that with negative values of $g_2$ the evolution breaks down very quickly, indicating a loss of hyperbolicity of the equations.

\section{Discussion}
\label{sec:discuss}

During inflation, higher derivative terms in the gravitational action may play a significant role. In this work we have studied the impact of four-derivative corrections to the scalar-tensor action on the dynamics of large perturbations during inflation, in the case where the inflaton is also the additional scalar degree of freedom of the modified theory. This builds on newly developed stable formulations for evolving Horndeski theories in NR \cite{Kovacs:2020pns, Kovacs:2020ywu,AresteSalo:2022hua,AresteSalo:2023mmd,AresteSalo:2023hcp,East:2020hgw,East:2021bqk}, and the broader development of NR tools for cosmology \cite{Aurrekoetxea:2024mdy}.

Modifications of this form to the inflaton action are generically harmful to slow roll dynamics. Even in the homogeneous and isotropic case, in the absence of any perturbations, they tend to change the shape of the effective potential in a way that prevents inflation ending, disrupts the slow roll dynamics by increasing the slope of the potential, or creates instabilities in the perturbations that would conflict with CMB observations. 

For this work, we restrict ourselves to cases in which inflation does still occur in the homogeneous limit, and that start within the regime of validity of the effective theory. We target the question of whether the non-linear dynamics of the perturbations we introduce are significantly changed by the higher derivative terms. In particular, we are interested in whether dynamical effects can drive us out of the regime of validity of the EFT. Broadly the conclusion is that they are not significantly changed, and would be well described by the dynamics observed in previous works in standard GR evolutions \cite{Corman:2022alv,East:2015ggf,Clough:2016ymm,Clough:2017efm}, with the potential modified to match the effective potential of the homogeneous and isotropic limit of the 4dST theory. Initial data that starts within the EFT regime tends to stay there, since GR dynamics initially dominate and the resulting expansion reduces the effect of the higher derivative terms over time.
 
Two caveats to this conclusion are as follows. First, we show that, in principle, it is possible to dynamically drive the field out of the weak-coupling regime from a starting point well within it. However, to do so one has to finely tune the setup, so such cases are unlikely to occur generically. Secondly, we show that the quadratic kinetic or ``$g_2$'' term in the action does have a significant effect, and seems to reduce the robustness of the model to perturbations. Although the contribution of this term is subdominant at each time (as required by weak coupling), its integrated effects result in a change to the final amplitude of the perturbations. This amplitude is then `frozen out' until the modes re-enter the horizon after inflation, where they could (in principle) have measurable effects.

This work provides a basis for the study of less restricted models in future, for example those in which the inflaton and scalar degree of freedom are independent, and where the formation of black holes may give rise to scalarisation. The initial conditions could also be made more general by an improved initial data solver, in which conformal flatness is not required.
One could also take advantage of recent developments in providing a well-posed initial value formulation for a more general class of theories \cite{Figueras:2024bba,Figueras:2025gal}.

\section{\label{sec:acknowledge}Acknowledgements}

We would like to thank Llibert Aresté Saló, Josu Aurrekoetxea, Maxence Corman and Will East for helpful discussions, and the entire \texttt{GRTL} Collaboration\footnote{\texttt{www.grtlcollaboration.org}} for their support and code development work. PF would like to thank the Enrico Fermi Institute and the Department of Physics of the University of Chicago for hospitality during the final stages of this work. KC is supported by an STFC Ernest Rutherford fellowship, project reference ST/V003240/1. PF, ADK and KC are supported by an STFC Research Grant ST/X000931/1 (Astronomy at Queen Mary 2023-2026). SB is supported by a QMUL Principal studentship.
This work used the DiRAC@Durham facility managed by the Institute for Computational Cosmology on behalf of the STFC DiRAC HPC Facility (www.dirac.ac.uk). The equipment was funded by BEIS capital funding via STFC capital grants ST/P002293/1, ST/R002371/1 and ST/S002502/1, Durham University and STFC operations grant ST/R000832/1. DiRAC is part of the National e-Infrastructure.
Calculations were also performed using the Sulis Tier 2 HPC platform hosted by the Scientific Computing Research Technology Platform at the University of Warwick. Sulis is funded by EPSRC Grant EP/T022108/1 and the HPC Midlands+ consortium. This research also utilised Queen Mary’s Apocrita HPC facility, supported
by QMUL Research-IT.
For the purpose of Open Access, the author has applied a CC BY public copyright licence to any Author Accepted Manuscript version arising from this submission.

\appendix

\section{Numerical methods and validation}
\label{sec:appendix}

 \begin{figure}[!ht]    
 \begin{center}
 \includegraphics[width=0.6\textwidth]{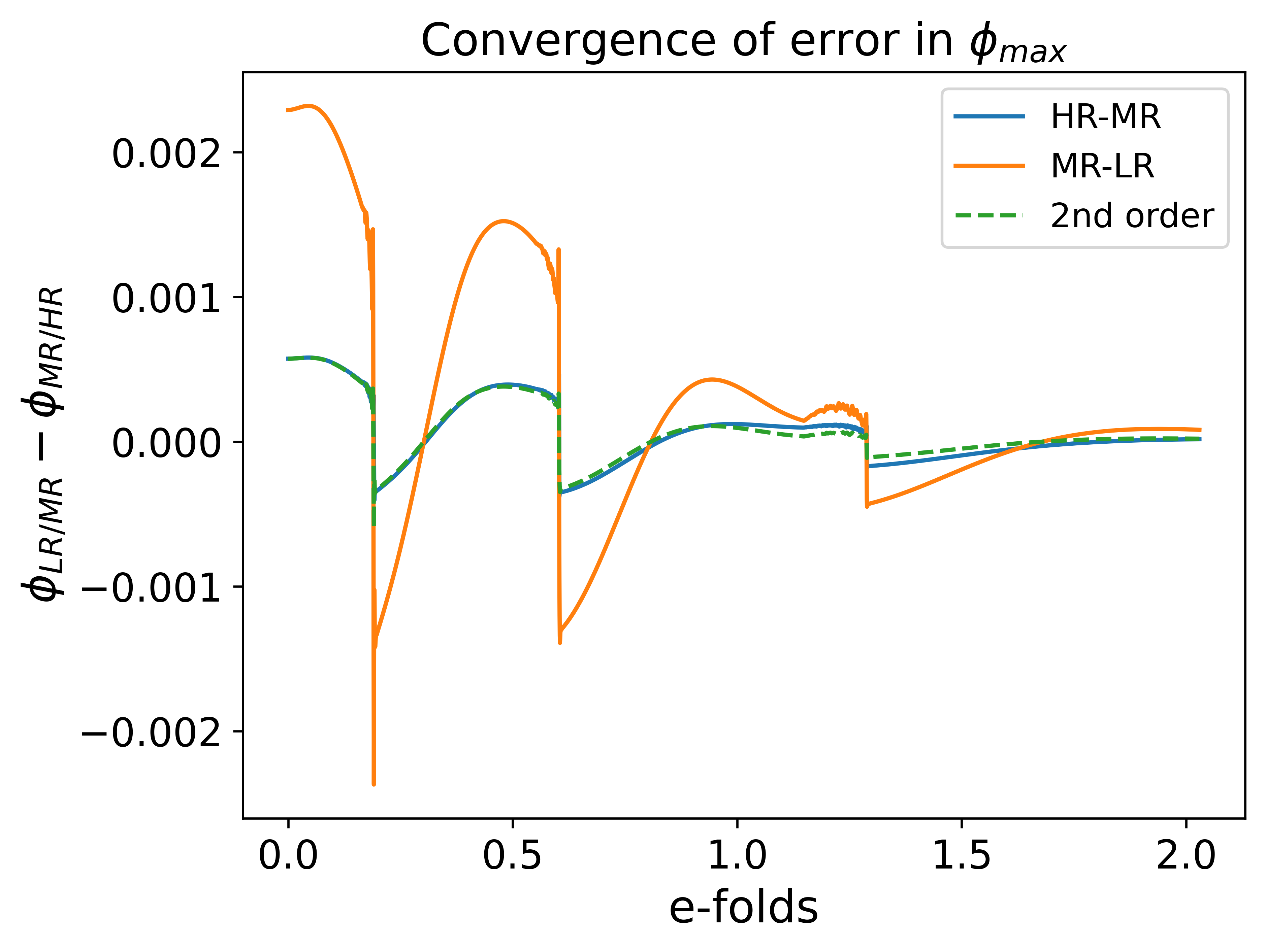}
    \caption{Convergence tests for $\phi_{max}$. We compare the differences in $\phi_{max}$ for low, medium, and high resolution simulations. This shows our results converge to 2nd order, appearing to be dominated by the error in the initial data.}
    \label{fig:convergence}
    \end{center}
\end{figure}

\begin{figure}[!ht]
    \begin{center}
    \includegraphics[width=0.6\textwidth]{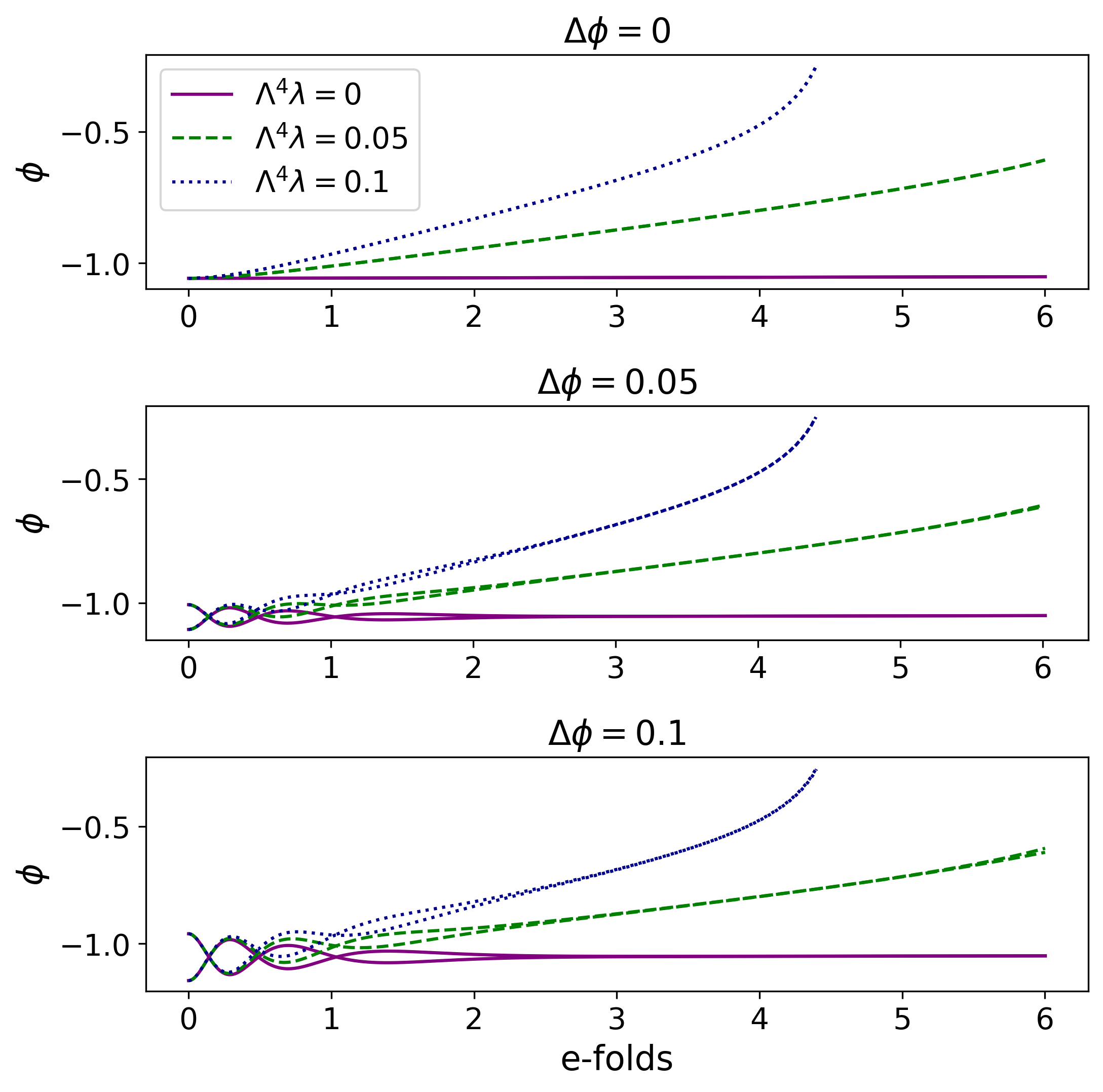}
     \caption{Plots of the maximum and minimum values of $\phi$ against N-folds of expansion, for nine values of the coupling $\lambda$ and the perturbation size $\Delta\phi$. These small perturbations in $\phi$ are expected to rapidly decay, as observed in all cases.}
     \label{fig:subhorizon_plots}
     \end{center}
\end{figure}

We evolve the coupled system of Klein-Gordon and general relativity equations using the standard CCZ4 formulation of \cite{Alic:2011gg} and the moving puncture gauge \cite{Bona:1994dr,Baker:2005vv,Campanelli:2005dd,vanMeter:2006vi} within the numerical-relativity code \textsc{grchombo} \cite{Andrade:2021rbd,Radia:2021smk,Clough:2015sqa}, which uses the method of lines, with an RK4 time integration and 4th order finite difference stencils for calculating spatial gradients. The initial data is generated using the code GRTresna \cite{Aurrekoetxea:2025kmm} using the CTTK method of \cite{Aurrekoetxea:2019fhr}.

During the simulations we monitor the dynamics of the scalar field by considering the following contributions to $\frac{\partial^2\phi}{\partial t^2}$, which we will illustrate in Fig \ref{fig:fine_tuned_plots}: \\
\begin{itemize}
    \item Friction: $\alpha K\Pi$ 
    \item Gradient: $\alpha D^i D_i \phi + D^i\phi D_i \alpha$
    \item Gauss-Bonnet: $\alpha \lambda'(\phi) \mathcal{L}^{GB}$
    \item Potential: $\alpha \frac{dV}{d\phi}$
\end{itemize}

We perform convergence tests to validate our simulations by repeating the runs at three different resolutions, with no additional levels of refinement and a number of grid points in the coarsest level $N = \left\{64,\, 128,\, 256\right\}$. We define the error as the difference in $\phi_{max}$ between two different resolutions. We expect the errors in our simulations to decrease at a predictable rate as we increase the resolution of the runs. We define the convergence factor,
\begin{equation}
    c(t) = \frac{|\phi_{\Delta_1} - \phi_{\Delta_2} |}{|\phi_{\Delta_2} - \phi_{\Delta_3}|} ~,
\end{equation}
where $\Delta_1,\Delta_2,\Delta_3$ are the spatial step sizes at the coarsest grid level for each resolution, from lowest resolution to highest. In the $\Delta \to 0$ limit,
\begin{equation}
    \lim_{\Delta \to 0}{c(t)} = \frac{\Delta_1^n - \Delta_2^n}{\Delta_2^n - \Delta_3^n}\,,
\end{equation}
where $n$ is the order of convergence. The top and bottom panels of Fig. \ref{fig:convergence} show the error in the maximum value of the scalar field for different resolutions in a typical simulation. The results confirm the expected approximate 2nd-4th order convergence.

We also check the behaviour matches that expected in the homogeneous case in the limit of small perturbations. In Figure \ref{fig:subhorizon_plots} the average value of the scalar field in its initial configuration (given in Eq. \eqref{eq:perturbations_form}) is fixed as the EsGB coupling $\lambda$ and the perturbation size $\Delta\phi$ are varied. The average initial field value $\phi_0$ is that which gives 100 e-folds of inflation in FLRW spacetimes with no higher derivative terms. It is expected that in these cases the small perturbations will die away, and that the number of efolds will be consistent with the homogeneous case.  with $\phi_0=-1.0574$ and $\Lambda^4\lambda$ values of 0, 0.05 and 0.1 giving 100, 7.2 and 4.5 e-folds of inflation respectively. 
This supports the wider validity of our numerical scheme, and shows that in the nearly-homogeneous limit the evolution is stable for these values of the coupling $\lambda$.
%\end{multicols}
\newpage
\printbibliography[title={Bibliography}]
\end{document}